\documentstyle[twocolumn,epsf]{jpsj}
\def \virg{\;\;,}
\def \point{\;\,.}
\def \kf{k_{\rm F}}
\def \vf{v_{\rm F}}
\def \e{{\rm e }}
\def \i{{\rm i }}
\def \d{{\rm d}}
\def \td{t_{\rm d}}

\def \Vc{V_{\rm c}}
\def \V2c{V_{\rm 2c}}
\def \sgn{{\rm sgn}}
\def \adj{{\rm adj}}
\def\ggs{\buildrel\textstyle > \over {\hbox{\raise0.2ex\hbox{$\sim$}}}}
\def\lls{\buildrel\textstyle < \over {\hbox{\raise0.2ex\hbox{$\sim$}}}}
\def\gsim{\,\lower0.75ex\hbox{$\ggs$}\,}
\def\lsim{\,\lower0.75ex\hbox{$\lls$}\,}
\def\snearrow{\mbox{\scriptsize$\nearrow$}}

\def\sswarrow{\mbox{\scriptsize$\swarrow$}}
\def\bnearrow{\mbox{\large$\nearrow$}}

\def\bswarrow{\mbox{\large$\swarrow$}}
\def\suparrow{\mbox{\scriptsize$\uparrow$}}
\def\sdownarrow{\mbox{\scriptsize$\downarrow$}}
\def\buparrow{\mbox{\Large$\uparrow$}}
\def\bdownarrow{\mbox{\Large$\downarrow$}}
\def\sleftarrow{\mbox{\scriptsize$\leftarrow$}}
\def\srightarrow{\mbox{\scriptsize$\rightarrow$}}
\def\runtitle{
Effects of Charge Ordering on Spin-Wave of 
 Quarter-Filled SDW States  
  }
\def\runauthor
{
 Yuh {\sc Tomio}, 
 Yasunari {\sc Kurihara}
 and   
 Yoshikazu {\sc Suzumura} 
}

\renewcommand{\theequation}{\arabic{section}.\arabic{equation}}

\title{   
Effects of Charge Ordering on Spin-Wave of 
 Quarter-Filled \\
Spin-Density-Wave States  
}

\author{
 Yuh {\sc Tomio}$^{1,}$%
\footnote{E-mail: tomio@edu2.phys.nagoya-u.ac.jp}%
,
 Yasunari {\sc Kurihara}$^{2,}$%
\footnote{E-mail: kurihara@es.htokai.ac.jp}
   and Yoshikazu  {\sc Suzumura}$^{1,}$%
\footnote{E-mail: e43428a@nucc.cc.nagoya-u.ac.jp}
}

\inst{
$^{1}$Department of Physics, Nagoya University, Nagoya 464-8602 \\ 
$^{2}$Research Institute for Higher Education Programs, 
 Hokkaido Tokai University, Sapporo 005-8601 \\ 
}

\recdate{January 15, 2002}

\abst
{
Spin-wave excitations  of    
   quarter-filled spin-density-wave  state, which  
    coexists with   charge  ordering,   have been studied 
  for   one-dimensional extended Hubbard model with the nearest-neighbor 
 repulsive interaction ($V$)  and   next-nearest-neighbor  one ($V_2$).  
  We calculate  dispersion relations 
 for the acoustic and optical spin-wave 
  within the random phase approximation.
 Our numerical calculation shows that energy spectrum of 
  the acoustic branch is well described by  a simple sine function form. 
 In the states  coexisting  with  charge-density-wave,  
   the spin-wave velocity decreases with increasing $V$ or $V_2$. 
  Our  numerical result, that velocity reduces to zero  
    in the limit of large $V$ or $V_2$,  
   is analyzed in terms of a  spin 1/2 Heisenberg model   
     with effective antiferromagnetic exchange interaction. 
}

\kword
{
 collective mode, spin-wave, 
  spin-density-wave, charge ordering, quarter-filled band, 
 extended Hubbard model, nearest-neighbor interaction, 
next-nearest-neighbor interaction
  }

\begin{document}
\sloppy
\maketitle
\section{Introduction}
\setcounter{equation}{0}
Organic conductors, 
 tetramethyltetraselenafulvalene (TMTSF) and 
  tetramethyltetrathiafulvalene (TMTTF) salts,  
\cite{Jerome,Yamaji} 
  exhibit, due to  an interplay of interaction and  low dimension, 
 exotic  spin-density-wave (SDW) states, which     coexist 
   with the  charge-density-wave (CDW), {\it e.g.},      
 4$\kf$~CDW in TMTTF salt  and   2$\kf$~CDW in TMTSF salt
($\kf$ denotes a Fermi wave  number). 
\cite{Pouget,Kagoshima} 
 These states have been analyzed  in terms of    
     the extended Hubbard model with several repulsive interactions.   
  Applying   the mean-field theory, it is shown that  
    the interactions with  finite-range  play a crucial role 
    for the  coexistence. 
 The  coexistence of  SDW  with 4$\kf$~CDW (2$\kf$~CDW) appears when  
   the nearest-neighbor repulsive interaction $V$,   
   (the next-nearest-neighbor repulsive interaction $V_2$) 
     becomes large.
\cite{Seo,Kobayashi} 

 The fluctuations around the SDW ground state have been 
  examined extensively by calculating 
   the collective modes 
     within the random phase approximation (RPA). 
 The modes for charge and spin fluctuations 
  in the presence of only  on-site repulsive interaction ($U$) 
   were evaluated for the incommensurate SDW 
\cite{Kurihara82,Takada84,Kurihara84,Psaltakis84,Maki87}  
 and for  the commensurate SDW state at quarter-filling,
\cite{Suzumura_JPSJ95,TanemuraPTP}
  where the latter  may be  relevant to above organic conductors.    
The commensurability energy induces   an excitation gap 
 in the charge excitation where the gap   becomes zero 
 at  $V = \Vc$. 
\cite{Suzumura97} 
 When  $V = \Vc$, 
    a coexistent state of  2$\kf$~SDW and 4$\kf$~CDW appears  
\cite{Seo} 
 and   the harmonic potential for  the charge fluctuation vanishes. 
 The presence of  $V_2$ leads to     
  an unusual  ground state where  2$\kf$~SDW coexists with  2$\kf$~CDW
 and also with  4$\kf$~SDW 
  for $V_2 >  \V2c$.
\cite{Kobayashi,TomioJPSJ} 
 When  $V_2 = \V2c$, one finds   
 the disappearance of  the  harmonic potential with respect to 
 the relative motion   between the density wave of the up spin 
    and that of the down spin. 
 For  large $V_2$,     
   a new collective mode with an excitation  gap appears 
  where the mode  describes the relative motion of 
    density waves with opposite spin 
    followed by  amplitude mode of 2$\kf$~CDW.  
 The excitation gap  vanishes at $V_2=\V2c$.  
\cite{TomioLET} 

 While the  spin-wave modes at quarter-filling was  calculated  for 
   the conventional Hubbard model,  
\cite{Suzumura_JPSJ95,TanemuraPTP} 
 much of their problems is not known in the presence of $V$ and $V_2$.  
  The excitation with long-wavelength has been calculated 
 within the RPA  for the case of $V \neq 0$ and $V_2=0$, 
\cite{Mori}
 and for the spin-wave velocity 
   as a function of $V$ or $V_2$ 
\cite{TomioPRB}
based on a functional integral formulation.  
\cite{Sengupta} 
The latter has shown that 
the spin-wave velocity decreases to zero (a finite value) 
with increasing $V$ ($V_2$) 
 when 2$\kf$~SDW coexists with 4$\kf$~CDW 
 (2$\kf$~SDW coexists with  2$\kf$~CDW and 4$\kf$~SDW). 
 In  the path integral method, it remains an open question how to 
 treat   the coupling to long-wavelength  spin fluctuations.
\cite{TomioPRB}   

In the present  paper,  we examine  the acoustic and optical 
 spin-wave  modes 
 in the presence of $V$ and $V_2$ (leading to the charge ordering) 
 for understanding  the properties of the several kinds of 
  transverse spin fluctuations 
  of the SDW ordered states coexisting  with  CDW.  
We calculate not only  the energy dispersion relation of them but also 
 their collective operators within RPA.   
In \S2, 
 the mean-field ground state and the response function are explained 
   for the calculation of  the spin-wave modes.  
In \S3, 
 the resulting energy spectrum of them is given in conjunction with 
 the characteristics of their collective coordinates including 
 several spin fluctuation.  In addition, the interaction dependence of 
the spin-wave velocities is shown 
    as a function of $V$ or $V_2$.   
  The effect of the charge ordering on the spin-wave   is examined 
 and is compared with that obtained by the path integral method.
\cite{TomioPRB} 
 In \S4, we 
 discuss  the effect of dimerization and 
 analyze  the behaviors of the spin-wave velocity 
   in the limit of large $V$ or $V_2$ 
 by using  an effective Hamiltonian of a spin 1/2 Heisenberg model.

\section{Mean-Field Ground State and Response Function}
\setcounter{equation}{0}
We study  a one-dimensional extended Hubbard model  
 at quarter-filling. The Hamiltonian is  given by 
\begin{eqnarray}
  \label{H}
	H   &=&  - \sum_{\sigma,j}  
         ( t - (-1)^j \td )
         ( C_{j\sigma}^\dagger  C_{j+1,\sigma} + {\rm h.c.} )
\nonumber \\
    & &	 {} +  H_{\rm {int}} \virg \\
  \label{Hint}
        H_{\rm {int}}  &=&   \sum_{j=1}^{N}
            (  U  n_{j \uparrow} n_{j \downarrow}
             + V    n_{j} n_{j+1}  
             + V_2  n_{j} n_{j+2})  \virg
\end{eqnarray}
where $C_{j\sigma}^{\dagger}$  
 denotes a creation  operator 
 of an electron at the $j$-th site with spin 
 $\sigma =(\uparrow, \downarrow)$, and satisfies a periodic 
 boundary condition 
 $C_{j+N,\sigma}^{\dagger}=C_{j \sigma}^{\dagger}$ 
   with the total number of lattice site $N$, 
 $n_{j}=n_{j \uparrow}+n_{j \downarrow}$ and 
 $n_{j\sigma}=C_{j\sigma}^{\dagger}C_{j\sigma}$.  
 The quantity $t$ and $\td$ are the energy of  the transfer integral 
 and  that of  the dimerization. 
 Quantities $U$, $V$ and $V_2$ correspond to coupling  constants 
 of repulsive interaction 
for the on-site, the nearest-neighbor site and 
 the next-nearest-neighbor site, respectively.  
 We take $t$  and the lattice constant as unity.  

 For a quarter-filled band, 
 the mean-fields (MFs) 
  of  SDW and CDW are  written as
 ($m=0,1,2$ and 3, and    
 $\sgn (\sigma) = +(-)$ for $\sigma = \uparrow (\downarrow)$) 
{\setcounter{enumi}{\value{equation}}
\addtocounter{enumi}{1}
\setcounter{equation}{0} 
\renewcommand{\theequation}{\arabic{section}.\theenumi\alph{equation}}
\begin{eqnarray}   
   S_{mQ_0} &=& \frac{1}{N} \sum_{\sigma=\uparrow,\downarrow}  
                \sum_{-\pi < k \leq \pi } \sgn (\sigma)  
                \left<C_{k\sigma}^\dagger C_{k+mQ_0,\sigma}
                 \right>_{\rm MF}  \virg 
                                  \label{OPS} \nonumber\\   \\ 
   D_{mQ_0} &=& \frac{1}{N} \sum_{\sigma=\uparrow,\downarrow}  
                \sum_{-\pi < k \leq \pi }   
                \left<C_{k\sigma}^\dagger C_{k+mQ_0,\sigma}
                \right>_{\rm MF} \virg    
                                   \label{OPD}  
\end{eqnarray} 
\setcounter{equation}{\value{enumi}}}%
 where $Q_0 \equiv 2\kf=\pi/2$ 
   with $\kf (=\pi/4)$ being the Fermi wave number.  
 The   $z$-axis is taken as   the quantized axis.   
 The expression $<O>_{\rm MF}$ denotes an average of  
  $O$ over  the MF Hamiltonian,
\cite{TomioJPSJ}  
   given by  
\begin{eqnarray}   \label{MFH}
   H_{\rm MF} & = & 
          \sum_{\sigma=\uparrow,\downarrow}   
         \sum_{-\pi< k \leq \pi }  
     \Bigl[ \Bigl( \varepsilon_k + \frac{U}{4} + V + V_2 \Bigr)
         C_{k\sigma}^\dagger C_{k\sigma}  
 \nonumber  \\
        & &  {}
       +     \Bigl\{   \Bigl( 
 (\frac{U}{2} -2V_2) D_{Q_0} - \sgn(\sigma) \frac{U}{2} S_{Q_0}
             \Bigr) 
 \nonumber  \\
        & & \hspace{5mm} {} \times 
         C_{k+Q_0 ,\sigma}^\dagger  C_{k\sigma} + {\rm h.c.}  \Bigr\}
 \nonumber  \\
        & & {} 
       +
       \Bigl( 
(\frac{U}{2} -2V + 2V_2) D_{2Q_0}
    - \sgn(\sigma) \frac{U}{2} S_{2Q_0} 
 \nonumber  \\ 
        & & \hspace{5mm} {}
    - 2 \i \td \sin k \Bigr)
         C_{k\sigma}^{\dagger} C_{k+2Q_0,\sigma} \Bigr]   
 \nonumber  \\ 
        & & {}
       + 
       NU \Bigl[ - \frac{1}{16} -\frac{1}{2} \Bigl( 
       {|D_{Q_0}|}^2 - {|S_{Q_0}|}^2 \Bigr)  \nonumber \\
 && {} 
     - \frac{1}{4} \Bigl( D_{2Q_0}^{2} - S_{2Q_0}^{2} 
       \Bigr) \Bigr]
       + 
       NV  \Bigl(  -\frac{1}{4} + D^2_{2Q_0}  \Bigr) 
 \nonumber  \\
        & & {}
       + NV_2 \Bigl( -\frac{1}{4} + 2 {|D_{Q_0}|}^2 
       - D^2_{2Q_0} \Bigr)   \virg
\end{eqnarray} 
 where  $\varepsilon_k=-2t \cos k$. 
 The expressions of each MF are given  such that      
 $S_0=0$, $D_0=1/2$, $S_{Q_0}=S^*_{3Q_0} \equiv S_1 \e^{\i \theta}$, 
$D_{Q_0}=D^*_{3Q_0} \equiv D_1 \e^{\i (\theta-\pi/2)}$, 
$S_{2Q_0}=S^*_{2Q_0} \equiv S_2$ and  $D_{2Q_0}=D^*_{2Q_0} \equiv D_2$. 
 Quantities  $S_1(> 0)$, $D_1 (\geq 0)$, $S_2$ and $D_2$    
  denote  amplitudes for 
2$\kf$~SDW, 2$\kf$~CDW, 4$\kf$~SDW and 4$\kf$~CDW, respectively. 
Their values depend on the corresponding ground states. 
The quantity $\theta$ denotes  a phase of SDW. 
\cite{TomioJPSJ}

%
\begin{figure}[tb]
\begin{center}
 \vspace{2mm}
 \leavevmode
 \epsfysize=7.5cm\epsfbox{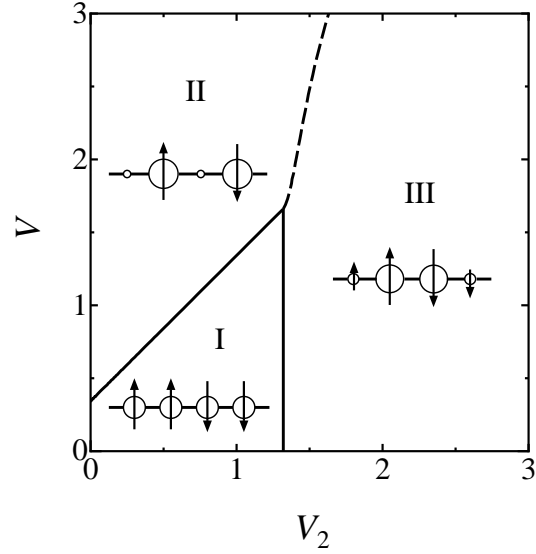}
 \vspace{-3mm}
\caption[]{
Phase diagram in the $V$-$V_2$ plane for $U=4$ and $\td=0$. 
\cite{TomioJPSJ}
Three kinds of  states correspond to 
a  pure state of 2$\kf$~SDW(I), 
a  state of  2$\kf$~SDW and 4$\kf$~CDW (II) and  
 a state of  2$\kf$~SDW, 2$\kf$~CDW and 4$\kf$~SDW (III). 
These states are  shown schematically in the region where  
 the arrow (circle) represents the spin (charge) density 
  at  each lattice site. 
The dashed curve denotes 
 a first-order transition  between regions II and III. 
 }
\end{center}
\end{figure} 
%
%
The ground state has been obtained previously
\cite{Seo,Kobayashi,TomioJPSJ}
where the phase diagram on the plane of $V_2$ and $V$ 
  is shown   in Fig.~1.
 When   $V_2=0$ and $\td=0$, 
 a  pure 2$\kf$~SDW state ($\theta=\pi/4$, $\phi=\pi/2$) (I) 
 is found      for $V < \Vc$ ($\simeq 0.34$)
  and  a coexistent state of 2$\kf$~SDW and 4$\kf$~CDW 
($\theta=0$, $\phi=\pi/2$) (II) 
 is found for $V > \Vc$ 
\cite{Seo,Suzumura97}, 
 where $\phi = \tan^{-1}(S_1/D_1)$. 
When $V=0$ and $\td=0$, 
 the  pure 2$\kf$~SDW state (I) changes to 
 a coexistent state of 2$\kf$~SDW, 2$\kf$~CDW and 4$\kf$~SDW 
 ($\theta=\pi/4$, $\phi < \pi/2$) (III) 
 for $V_2 > \V2c$ ($\simeq 1.32$).
\cite{TomioJPSJ}
Note that   the  finite-range interactions 
 $V$ and $V_2$ contribute to 
 the appearance of the quantities  
 $D_1$ and $D_2$.
 In the following, 
 we use  the states I,  II  and  III 
 for those  in the regions  I, II and III, respectively.

The spin-wave is examined for the above three ground states.
 The  spectrum of the spin-wave modes is calculated  
 from the pole of  the response function  given by 
\cite{Suzumura_JPSJ95}
\begin{eqnarray} \label{RR}
\label{nq0}
\stackrel{\leftrightarrow}{R} (q, \omega) &=& \frac{-1}{2N} 
          \int^{\beta}_{0} \hspace{-3mm} \d\tau 
   \left< T_{\tau} \Phi(q,\tau) \Phi^{\dagger}(q) \right>
       \e^{\i \omega_n \tau} 
      \big|_{\i \omega_n \rightarrow \omega+\i 0} 
 \virg \nonumber \\
\end{eqnarray}  
where $\omega_n (=2\pi n / \beta)$ and   
  $\beta^{-1}(= k_{\rm B}T)$ denote the Matsubara frequency 
    and temperature, respectively.
 The components of operator  $\Phi^{\dagger}(q)$ consist of 
  the transverse spin fluctuations given by 
$S_{\alpha m}(q)=\sum_{-\pi<k\leq\pi}
 \psi_{k}^\dagger \sigma_{\alpha} \psi_{k+q+mQ_0}$ 
 ($\alpha=x,y$) 
where $\psi_{k}^\dagger=(C^\dagger_{k\uparrow},C^\dagger_{k\downarrow})$
 and $\sigma_{\alpha}$ ($\alpha=x,y$) are Pauli matrices. 
For the state I ($\theta = \pi/4$) and 
  the  state  II  
 ($0 \leq  \theta <\pi/4$ due to $\td$),  
 the operator $\Phi^{\dagger}(q)$ can be taken as  
\cite{Suzumura_JPSJ95}
\begin{eqnarray} \label{Phi4}
   \Phi^{\dagger}(q) = 
 ( S_{x1}^\dagger (q),  S_{x3}^\dagger (q), 
   S_{y0}^\dagger(q), S_{y2}^\dagger (q) ) \point
\end{eqnarray}  
 We note that, for small $q$, 
  two components $S_{x1}^\dagger (q)$ and  
   $S_{x3}^\dagger (q)$ are related directly to the transverse spin 
 fluctuation of the SDW with wave numbers $Q_0$ and $-Q_0$ respectively 
 and that 
 other components of $S_{y0}^\dagger (q)$ and $S_{y2}^\dagger (q)$
 denote the fluctuations correlated with 
  those of the wave numbers 0 and $2Q_0$. 
The coupling to long-wavelength  spin fluctuation $S_{y0}^\dagger (q)$ 
 gives a renormalization of the spin-wave velocity 
 multiplied by a  factor $[1-UN(0)]^{1/2}$  in the weak coupling limit 
\cite{Poilblanc88}  
where $N(0)$ is the density of state per spin at Fermi level.  
For  the coexistent state of 2$\kf$~SDW, 2$\kf$~CDW and 4$\kf$~SDW 
(the state III) with the spin modulation with $Q_0$ and $2Q_0$, 
 one needs to extend  
  the operator  $\Phi^{\dagger}(q)$ into  that with  eight components 
 written as  
\begin{eqnarray} \label{Phi8}
   \Phi^{\dagger}(q) &=& 
 ( S_{x1}^\dagger (q),  S_{x3}^\dagger (q), 
   S_{x0}^\dagger(q), S_{x2}^\dagger (q), 
\nonumber \\
 & & 
   S_{y1}^\dagger (q),  S_{y3}^\dagger (q), 
   S_{y0}^\dagger(q), S_{y2}^\dagger (q) ) \point
\end{eqnarray}  
In addition to $S_{x1}^\dagger (q)$ and $S_{x3}^\dagger (q)$, 
 the component $S_{x2}^\dagger (q)$ for small $q$ is also   
  related directly to the transverse spin 
   fluctuation of the SDW with $2Q_0$  while  other components
 comes from  coupling to  these three components. 
Operators of eqs.~(\ref{Phi4}) and (\ref{Phi8}) act to create the 
transverse spin fluctuations by  raising or lowering  the spins 
 within the same sites (i.e.,  rotation of single electron due to 
 on-site interaction). 
 However,  the finite-range interactions induce
   the transverse spin fluctuations with the  
  spin current (i.e., spin rotation followed by 
    neighboring electrons), 
   which is related  to the bond spin-density-wave (BSDW) state. 
\cite{Japaridze95,MNakamura00} 
In the present paper, we neglect the effect of spin current,
 which  is left for the study in the separated  paper 
  together with  the problem of BSDW. 
 Based on  such a   consideration,    
  eq.~(\ref{RR}) within  RPA is calculated as 
\begin{eqnarray} \label{RRPA}
 \stackrel{\leftrightarrow}{R}^{\rm RPA} (q, \omega) &=& 
   \frac{\stackrel{\leftrightarrow}{\Pi}(q, \omega)}
   {1+U\stackrel{\leftrightarrow}{\Pi}(q, \omega)} \virg 
\end{eqnarray}
where the polarization function, 
  $\stackrel{\leftrightarrow}{\Pi} (q, \omega)$,
 is evaluated  in terms of the   MF 
 Hamiltonian,
  eq.~(\ref{MFH}).   
\begin{eqnarray}
\label{RRMF}
\stackrel{\leftrightarrow}{\Pi} (q, \omega) &=& \frac{-1}{2N} 
          \int^{\beta}_{0} \hspace{-3mm} \d\tau 
   \left< T_{\tau} \Phi(q,\tau) \Phi^{\dagger}(q) \right>_{\rm MF}
\nonumber \\
   && \hspace{15mm} {} \times   \e^{\i \omega_n \tau} 
      \big|_{\i \omega_n \rightarrow \omega+\i 0} 
 \point 
\end{eqnarray}
The pole of 
 eq.~(\ref{RRPA})  determines  
 the excitation spectrum of the collective mode, $\omega_{\gamma}(q)$, 
 i.e.,  
\begin{eqnarray} \label{EQpole}
 \det \left(  
   1+U\stackrel{\leftrightarrow}{\Pi}(q, \omega_{\gamma}(q))
\right) &=& 0 \virg
\end{eqnarray}
where $\gamma=T1$ or $T2$ and $\omega_{T1}$ ($\omega_{T2}$) denotes 
 the acoustic (optical) mode.  
The eigenvector, $\tilde{\Phi}^{\gamma}(q)$, 
 which describes  the collective mode, 
 is obtained by  
\begin{eqnarray} \label{eigen}
 \left(  
   1+U\stackrel{\leftrightarrow}{\Pi}(q, \omega_\gamma(q))
\right) \tilde{\Phi}^\gamma(q) &=& 0 \point
\end{eqnarray}
Actually, in terms of $\Phi(q)$ and $\tilde{\Phi}^\gamma(q)$, 
the operator for the  spin-wave with $\omega_{\gamma}(q)$ 
is  written   as 
\begin{eqnarray} \label{eta}
 \eta_\gamma(q) &=& \tilde{\Phi}^{\gamma\dagger}(q)\Phi(q)
\virg
\end{eqnarray}
  an explicit form  of which will be given  in the next section. 
 For   $\omega \simeq \omega_\gamma(q)$,
 eq.~(\ref{RRPA}) is rewritten as 
\cite{TanemuraPTP} 
\begin{eqnarray}
 \label{Rres}
   \stackrel{\leftrightarrow}{R} ^{\rm RPA}(q, \omega)  &\simeq& 
   \frac{S_1^2}{2} 
   \frac{\stackrel{\leftrightarrow}{A^\gamma}(q,\omega_{\gamma}(q))}
        {\omega^2-\omega_\gamma^2(q)} \virg \\
 \label{A}
   \stackrel{\leftrightarrow}{A^\gamma}(q,\omega_\gamma(q)) &=& 
 \frac{2}{S_1^2}
 \frac{\adj\left(1+U\stackrel{\leftrightarrow}{\Pi}(q,\omega_\gamma(q))\right)}
 {\frac{\partial}{\partial(\omega^2)} 
 \left(\det(1+U\stackrel{\leftrightarrow}{\Pi}
 (q,\omega))\right)}_{\omega \rightarrow \omega_\gamma(q)} 
\hspace{-3mm}, \nonumber \\
\end{eqnarray}
where the matrix 
 $\stackrel{\leftrightarrow}{A^\gamma}(q,\omega_\gamma(q))$ is  
proportional to the residue and 
$\adj\left(1+U\stackrel{\leftrightarrow}{\Pi}(q,\omega_\gamma(q))\right)$
 is  the adjoint matrix.  
The spectral weight of the collective mode 
 is examined by calculating 
    following diagonal element, 
\begin{eqnarray} 
\label{Am}
  A_{m}^\gamma(q) &\equiv& \left|
    \left( 
\stackrel{\leftrightarrow}{A^\gamma}(q,\omega(q)) 
    \right)_{mm}
 / U \right|    \!\virg \hspace{2mm} (m=1,3,0,2) \point \nonumber \\
\end{eqnarray}
 When $\td=0$ and $V=V_2=0$, one finds that,
   in the limit of small $U$, 
 $A^{T1}_{1}(0)=A^{T1}_{3}(0)=2\pi\vf$ with $\vf=\sqrt{2}$ and 
 $A^{T2}_{1}(0)=A^{T2}_{3}(0)=A^{\gamma}_{0}(0)=A^{\gamma}_{2}(0)=0$. 

We     examine 
 the $V$ (or $V_2$)  dependence of 
 the velocity 
 of   acoustic spin-wave given by  
\begin{eqnarray} \label{velocity}
 v = \lim_{q \rightarrow 0} \frac{\omega_{T1}(q)}{q} \virg
\end{eqnarray}
which is compared with the previous work.
\cite{TomioPRB}

\section{Spin-Wave Modes}
\setcounter{equation}{0}

 The characteristic properties  of the spin-wave modes 
 are examined for the various  kinds of SDW  ground states, which 
 are given by the states  I, II and III.
 The numerical calculation is performed   by taking $U=4$ and 
 $t = 1$ where  $\td = 0$  in this section. 

\subsection{Effect of $V$}

%
First, we calculate the spin-wave spectrum $\omega_\gamma(q)$  
as a function of $V$  for $V_2=0$.  
 In Fig.~2,   the spectrum of both 
  the acoustic and optical spin-wave modes
    obtained from eq.~(\ref{EQpole})  
     are shown for  $V=0$, 0.5, 1.5 and 2.0 
         with  $V_2=0$ and $\td=0$ 
 where the state II 
  is obtained for   $V > \Vc$ ($\simeq 0.34$).  
Our numerical calculation within the visible scale 
 shows that spectrum of the acoustic mode 
  takes a simple sine function form with respect to  the wave number $q$, 
 i.e.,  $\omega_{T1}(q)=\omega_{T1}(Q_0/2)\sin (2q)$. 
The optical mode,  $\omega_{T2}(q)$, is well separated from the continuum  
 for the parameters in Fig.~2.
 The spectrum  of the state  II is similar to that of the state  I.
%
%
\begin{figure}[tb]
\begin{center}
 \vspace{2mm}
 \leavevmode
 \epsfysize=7.5cm\epsfbox{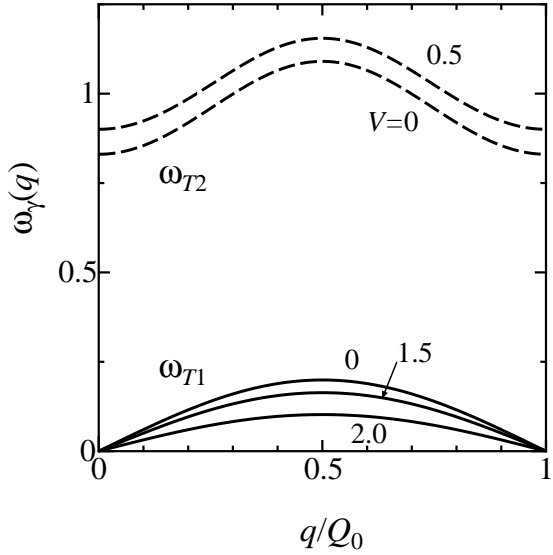}
 \vspace{-3mm}
\caption[]{
Excitation spectrum of the spin-wave modes, 
$\omega_{T1}(q)$ (acoustic mode) 
and $\omega_{T2}(q)$ (optical mode) for   $U=4$, $V_2=0$ and $\td=0$. 
 }
\end{center}
\end{figure} 
%

The  spectral weights for the acoustic mode,  
 $A_{m}^{T1}(q)$, 
are shown in Fig.~3  
 where  $A_{1}^{T1}(q)$ and $A_{3}^{T1}(q)$ 
 ($A_{0}^{T1}(q)$ and $A_{2}^{T1}(q)$) 
are weights 
 for spin fluctuations of  $x$ ($y$) direction  
 with wave numbers $Q_0+q$ and $-Q_0+q$ ($q$ and $2Q_0+q$), respectively. 
For the spectral weights  with $q \simeq 0$ or $Q_0$,   
the dominant weights are given by  the fluctuations  with 
  $ \simeq \pm Q_0$   
(note that $S_{y0}(Q_0)=S_{y1}(0)$ and $S_{y2}(Q_0)=S_{y3}(0)$), 
 as shown  for a  model with only $U$. 
\cite{TanemuraPTP}
Figure~3 indicates  that the deviation of the spin orientation 
 arising from the spin-wave 
 changes continuously  from the $x$ direction to the $y$ direction 
  when  $q$ increases from 0 to $Q_0$. 
 It is found that  all the weights are suppressed by $V$. 
%
\begin{figure}[tb]
\begin{center}
 \vspace{2mm}
 \leavevmode
 \epsfysize=7.8cm\epsfbox{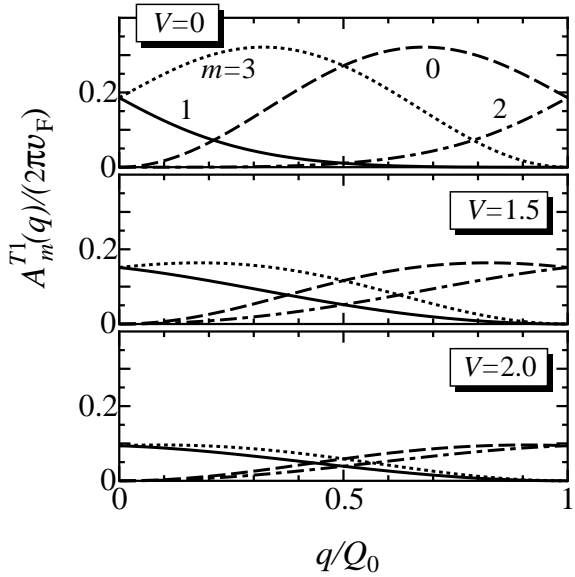}
 \vspace{-3mm}
\caption[]{
Normalized spectral weight of the acoustic mode, 
$A^{T1}_{m}(q)/(2\pi\vf)$, 
 for  $V=0$, 1.5 and 2.0 in Fig.~2.
 }
\end{center}
\end{figure} 
%

%
\begin{figure}[tb]
\begin{center}
 \vspace{2mm}
 \leavevmode 
\epsfxsize=7cm\epsfbox{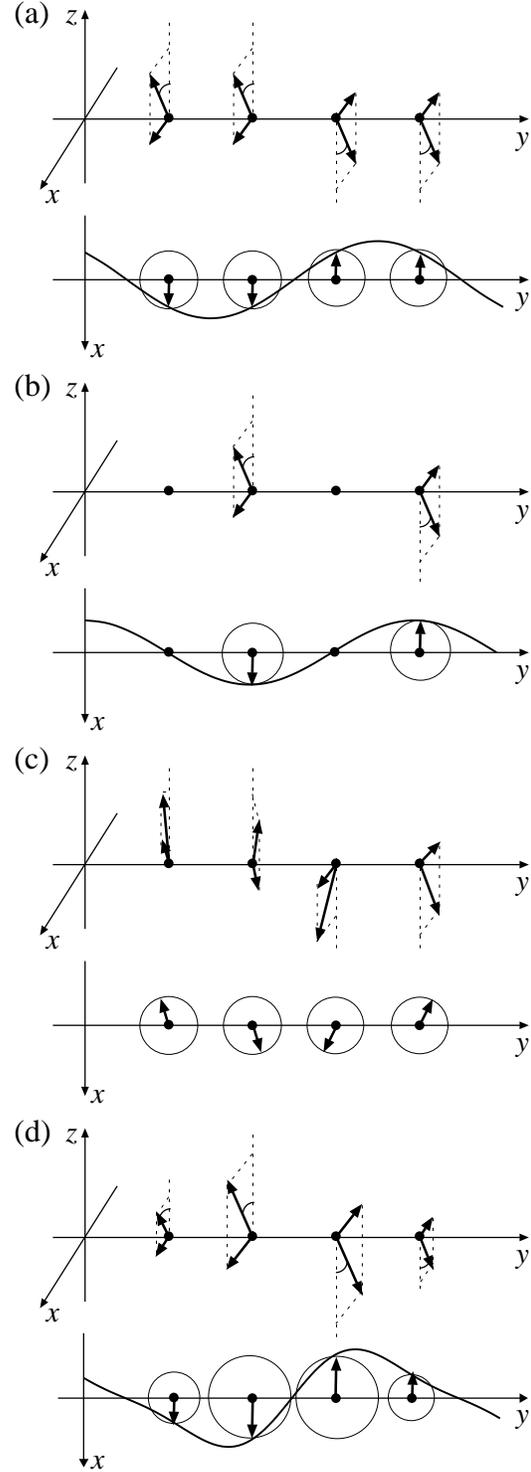}
 \vspace{-3mm}
\caption[]{
Schematics of spin-wave for $q \rightarrow 0$
 are shown. 
 (a) acoustic mode in the state I, 
 (b) acoustic mode in the state II, 
 (c) optical  mode in the state I and  
 (d) acoustic mode in the state III.  
The lower part of each figure represents 
 a projection of the spin on the $x$-$y$ plane. 
}
\end{center}
\end{figure}
%
Here we examine the behavior in the limit of small $q$. 
 From the numerical result of eq.~(\ref{eigen}), we found that   
 the eigenvector $\tilde{\Phi}^{T1}(q)$ 
    for the acoustic mode  is written as 
$
\tilde{\Phi}^{T1}(q \rightarrow 0)
 \propto
(\e^{\i \theta}, \e^{-\i \theta}, 0, 0) 
$
 where $\theta=\pi/4$ for $V < \Vc$ and 
  $\theta=0$ for $V > \Vc$. 
Then, the spin-wave operator of eq.~(\ref{eta}) in the limit of 
 $q \rightarrow 0$ may be  expressed as 
\begin{eqnarray} \label{eta2T1}
 \eta_{T1}(0) &\propto& \sum_{j} S^x_{j} \cos (Q_0 r_j + \theta)
\virg
\end{eqnarray}
where 
$S^\alpha_j= \psi^\dagger_{j} \sigma_{\alpha} \psi_{j}/2$ 
 and 
$
\psi^\dagger_{j}=(C^{\dagger}_{j\uparrow},C^{\dagger}_{j\downarrow}) 
$. 
 Equation (\ref{eta2T1}) denotes  that the gapless excitation at $q=0$ 
 is described by rotating  the spin quantization axis 
toward $x$ direction  uniformly.
  The spatial dependence is   illustrated in Figs.~4(a) and 4(b)
    for $V < \Vc$  and for $V > \Vc$, respectively,
 which are written simply   as   
$(\nearrow \nearrow \swarrow \swarrow)$  for $V < \Vc$ 
 and $(\cdot \nearrow \cdot \swarrow)$ for $V > \Vc$. 
Similarly, $\omega_{T1}(Q_0)$, corresponds  to  
 the excitation with a uniform rotation  toward $y$ direction.

 Another spectrum $\omega_{T2}(q)$ of the optical mode   
 has a gap.   
 The  corresponding 
  eigenvector $\tilde{\Phi}^{T2}(q)$  is given by 
$
\tilde{\Phi}^{T2}(q \rightarrow 0)
 \propto
(\e^{\i \theta}, -\e^{-\i \theta}, 0, c) 
$
 where $\theta=\pi/4$ and $c > 0$ for $V < \Vc$, and 
  $\theta=0$ and $c=0$ for $V > \Vc$. 
Thus, it is found that the spin-wave operator for the optical mode 
 in the limit of  $q \rightarrow 0$ is given by  
\begin{eqnarray} \label{eta2T2}
 \eta_{T2}(0) &\propto& \sum_{j} 
\left[
 2 S^x_{j} \sin (Q_0 r_j + \theta)
+ \i c S^y_{j} \cos (2Q_0 r_j) 
\right] \point
\nonumber \\ 
\end{eqnarray}
For $V < \Vc$, 
the spatial dependence is shown  schematically as 
$
(\nwarrow \: \nearrow \: \searrow \: \swarrow)
$  
 where  
   a  plane of $(\nwarrow \: \nearrow)$ 
    is  not parallel to  
  that of $(\searrow \: \swarrow)$ 
   due to  $c \neq 0$.  
 The  explicit  spin configuration  is illustrated in Fig.~4(c). 
For $V > \Vc$, 
 it is given by 
$
(\sleftarrow \buparrow \srightarrow \bdownarrow)
$  
where the symbol $\sleftarrow$ denotes a component of $S_{j}^x$.  
In Fig.~5, the gap of  the optical mode, 
$\omega_{T2}(0)$, is shown as a function of $V$. 
The thin dotted curve denotes the bottom of the continuum.  
With increasing $V$, 
   $\omega_{T2}(0)$ moves into the continuum   
   at the location shown by the arrow, 
The inset shows the corresponding spectral weight. 
 The spectral weight, 
  $A_m^{T2}$ (m=1,3) 
  reduces and becomes  zero  
 at $V \simeq 1$ corresponding to the arrow,
  i.e.,  
   above which the optical mode disappears. 
 For  $q \rightarrow 0$, the optical mode  in the state I 
 contains the fluctuation of $S_{y2}(q)$ 
 in addition to   the fluctuations of $S_{x1}(q)$ and  $S_{x3}(q)$   
  while  $S_{y2}(q)$ is absent for the mode  in the state II.
%
\begin{figure}[tb]
\begin{center}
 \vspace{2mm}
 \leavevmode
 \epsfysize=7.5cm\epsfbox{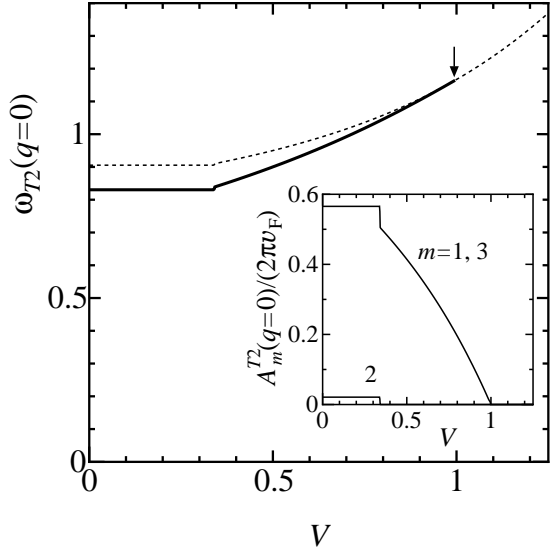}
 \vspace{-3mm}
\caption[]{
$V$ dependence of the gap, $\omega_{T2}(0)$, 
 for $U=4$, $V_2=0$ and $\td=0$ where 
$\omega_{T2}(q)$ disappears at the location shown by the arrow.
 The  dotted curve denotes the bottom of the continuum.  
In the  inset,  the corresponding spectral weight, 
$A^{T2}_{m}(0)/(2\pi\vf)$ is shown and 
 the  weight is zero if not shown explicitly. 
}
\end{center}
\end{figure}
%
%
\begin{figure}[tb]
\begin{center}
 \vspace{2mm}
 \leavevmode
 \epsfysize=7.5cm\epsfbox{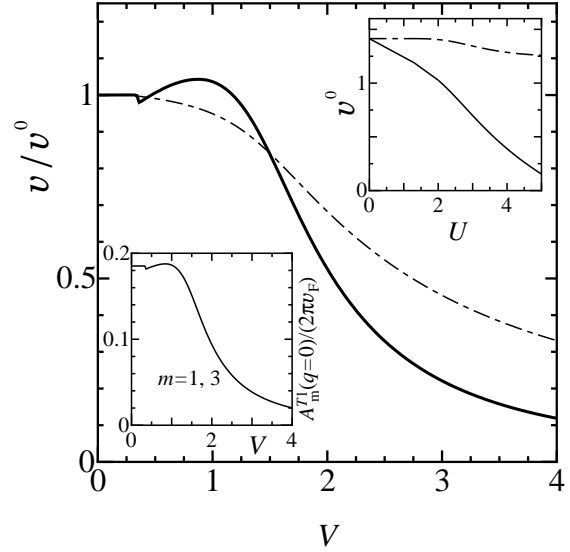}
 \vspace{-3mm}
\caption[]{
$V$ dependence of the spin-wave velocity $v$, 
for $U=4$, $V_2=0$ and $\td=0$ where 
 the  dot-dashed curve denotes the previous result obtained by 
a path integral method. 
\cite{TomioPRB}
 In the upper inset, $v^0$ as a function of $U$ 
 while,   
$A^{T1}_{m}(0)/(2\pi\vf)$ corresponding to the main figure 
 is shown  in the lower inset. 
}
\end{center}
\end{figure}
%
%

Now, we examine  the spin-wave velocity $v$, which is evaluated by using 
 eq.~(\ref{velocity}). 
 Figure 6 represents the $V$ dependence of $v$  
 where the dot-dashed curve denotes the result obtained 
 by the path integral method.
\cite{TomioPRB}
The quantity $v^0$, which 
  stands for the velocity in the case of $U=4$ and 
 $V=V_2=\td=0$, is given by 
   $v^0=0.39$ ($1.29$) for the solid (dot-dashed) curve. 
  In the upper inset,
 the $U$ dependence of $v^0$ is shown,
 where the solid curve and dot-dashed curve 
  correspond to those in the main figure. 
There is a small jump at $V=\Vc$ 
   due to the discontinuity of $S_1$ and $D_2$, which occurs  
 only for $\td=0$.
\cite{Seo}
With increasing  $V ( > \Vc)$, 
 $v$ (solid curve) takes a slight hump and 
   decreases to zero more rapidly  
 than  that of the dot-dashed curve. 
 The coupling to long-wavelength spin fluctuations was neglected 
 in  the path integral method. 
 Therefore, it is found  that 
 the velocity is strongly renormalized  by  
   the  coupling to such a  spin fluctuation. 
The hump is  not essential since it disappears 
     for  $\td \neq 0$ (see Fig.~10). 
The behavior of $v$ for large $V$ can be understood as 
 that of  spin 1/2 Heisenberg model with 
the effective exchange interaction $J \sim t^4/(UV^2)$, which  
 is   discussed in \S4. 
The lower inset  shows the corresponding weights of 
$A_{m}^{T1}(0)$ ($m=1,3$), which 
 decreases with increasing $V$ in a way   similar to  $v$.  

Here we comment on the  fact that 
 the spin-wave velocity as a function of $V$ 
    changes  rapidly for $V > \Vc$ while the velocity    
     remains the same as that of $V=0$ for $V < \Vc$. 
The latter  behavior of the velocity originates in 
 the present scheme of treating the response function. 
For the case of $V < \Vc$, in which the pure 2$\kf$~SDW state (I) is 
 realized and the MF is independent of the 
nearest-neighbor interaction $V$,   the quantities 
 $\stackrel{\leftrightarrow}{\Pi}(q,\omega)$ in eq.~(\ref{RRMF})
and  
$\stackrel{\leftrightarrow}{R}^{\lower1mm\hbox{\tiny RPA}}\!\!(q,\omega)$
 in eq.~(\ref{RRPA}) do not depend on $V$. 
 If we consider, however,  the transverse spin fluctuations with 
 the spin current which come from the term of the 
   nearest-neighbor interaction $V$, 
it is expected that the corresponding response function 
 $\stackrel{\leftrightarrow}{R}^{\lower1mm\hbox{\tiny RPA}}\!\!(q,\omega)$ 
 explicitly includes the interaction $V$  
  though the extended polarization function 
 $\stackrel{\leftrightarrow}{\Pi}(q,\omega)$  
is still independent of $V$. 
 Thus, the $V$ dependence of the spin-wave velocity may 
 appear in the state I.

\subsection{Effect of $V_2$}

Next, we calculate  the spin-wave mode as a function of 
 $V_2$ in the state  III   
 by using eqs.~(\ref{Phi8}) and (\ref{EQpole}). 
Figure 7 shows  
 the excitation spectrum of 
  the acoustic mode $\omega_{T1}(q)$  with some choices of $V_2$   
  for  $U=4$, $V=0$ and $\td=0$, 
    where    
  the coexistent state of 
    2$\kf$~SDW, 2$\kf$~CDW and 4$\kf$~SDW  is 
       found   for $V_2 > \V2c$ ($\simeq 1.32$). 
 The optical mode $\omega_{T2}(q)$  
   moves  into the continuum  
    leading to a disappearance of the mode for  $V_2 (\gsim 1.35 )$   
      being just above $\V2c$.
 With increasing $V_2 ( > \V2c)$, the spectrum $\omega_{T1}(q)$ 
   decreases.  
We note that the dispersions  for $V_2=1.5$ and 2.0 
  is also well described by a relation,   
 $\omega_{T1}(q)=\omega_{T1}(Q_0/2)\sin (2q)$. 
%
%
\begin{figure}[tb]
\begin{center}
 \vspace{2mm}
 \leavevmode
 \epsfysize=7.5cm\epsfbox{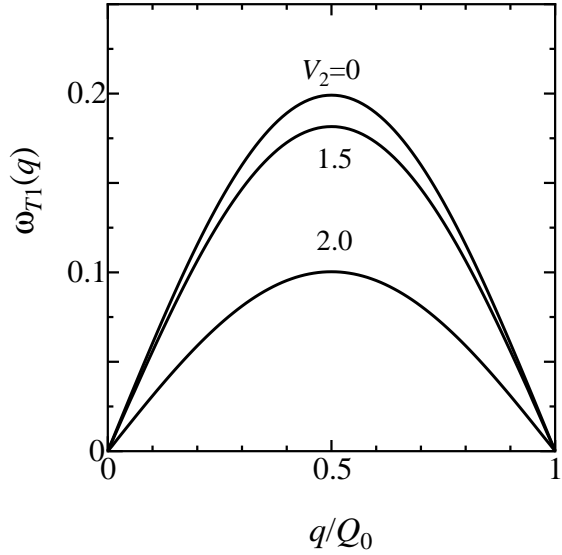}
 \vspace{-3mm}
\caption[]{
Excitation spectrum of the spin-wave mode, 
$\omega_{T1}(q)$, with some choices of $V_2$ 
 for  $U=4$, $V=0$ and $\td=0$. 
}
\end{center}
\end{figure}
%
\begin{figure}[tb]
\begin{center}
 \vspace{2mm}
 \leavevmode
 \epsfysize=7.5cm\epsfbox{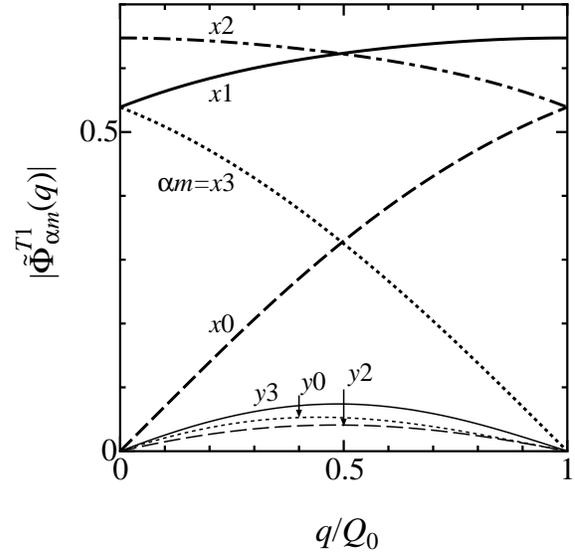}
 \vspace{-3mm}
\caption[]{
$q$ dependence of the elements of 
 eigenvector $|\tilde{\Phi}^{T1}(q)|$, corresponding 
to the case of $V_2=2.0$ in Fig.~7. 
}
\end{center}
\end{figure}
%

The corresponding eigenvector $\tilde{\Phi}^{T1}(q)$ 
  for  $V_2=2.0$ is shown in Fig.~8. 
For the calculation of  the eigenvector in eq.~(\ref{eigen}),
 we choose  
  seven components except for $S_{y1}(q)$  
    in  $\Phi(q)$ of eq.~(\ref{Phi8}) 
 since there are two degenerate modes, i.e.,  
 the mode with dominant contribution from $S_{xm}(q)$ and 
 the mode with that from  $S_{ym}(q)$. 
 The element $|\tilde{\Phi}^{T1}_{y1}(q)|$ corresponding to 
 $S_{y1}(q)$ is negligibly small, and then 
 the spectrum calculated from seven components 
  is actually nearly  the same as that of Fig.~7. 
 It is worthy to note that 
  the mode with  small $q$ 
 in  the region III of Fig.~1    is  composed  of 
not only $S_{xm}(q)$ ($m=1,3$) but also $S_{x2}(q)$ (Fig.~8).
Such a feature can be understood by considering 
 the spatial configuration, 
 in which 
 the spin modulation is of type 
$
(\suparrow \buparrow \bdownarrow \sdownarrow) 
$
 (Fig.~1)  and  the variation of spin moment 
 comes from the presence of 4$\kf$~SDW. 
 In this case,  the spin-wave operator expressing 
 the gapless excitation at $q \rightarrow 0$  
   is  expected to take a  form of  
\begin{eqnarray} \label{eta3}
 \eta_{T1}(0) &=& \sum_j  S^x_{j} 
\left[
 2B_1 \cos (Q_0 r_j + \theta) + B_2 \cos (2Q_0 r_j)
\right] \virg \nonumber \\
\end{eqnarray}
 which means a uniform rotation   toward $x$ direction 
 with keeping the relative angle between the each spin,  
 like   
$
(\snearrow \bnearrow \bswarrow \sswarrow) 
$. 
Here, the coefficients $B_1$ and $B_2$ satisfies
   the relation $B_2/B_1=S_2/S_1$  
 according to the spin modulation in the ground state. 
 In eq.~(\ref{eta3}), 
  the term of $B_2$ comes from   the fluctuation of $S_{x2}(q)$. 
Actually, our numerical calculation shows that 
\begin{eqnarray} \label{Phi7}
 \tilde{\Phi}^{T1}(q)\big|_{q \rightarrow 0} &=& 
\left(
 \tilde{\Phi}^{T1}_{x1}(q), \tilde{\Phi}^{T1}_{x3}(q), 
 \tilde{\Phi}^{T1}_{x0}(q), \tilde{\Phi}^{T1}_{x2}(q), \right. 
\nonumber \\
 & & {}\left.
 \tilde{\Phi}^{T1}_{y3}(q), \tilde{\Phi}^{T1}_{y0}(q),
 \tilde{\Phi}^{T1}_{y2}(q)
\right) \Big|_{q \rightarrow 0}
\nonumber \\
 &\propto& {} 
\left(
 \e^{\i \theta}, \: \e^{-\i \theta}, \: 0, \: c', \: 0, \: 0, \: 0 
\right) \virg
\end{eqnarray}
where $\theta=\pi/4$ and 
$
c'=|\tilde{\Phi}^{T1}_{x2}(0)|/|\tilde{\Phi}^{T1}_{x1}(0)|
  =S_2/S_1 
$ 
($\simeq 1.2$ for $V_2=2.0$).  
 The spatial dependence of  
this spin-wave  is explicitly illustrated in Fig.~4(d). 
%
\begin{figure}[tb]
\begin{center}
 \vspace{2mm}
 \leavevmode
 \epsfysize=7.5cm\epsfbox{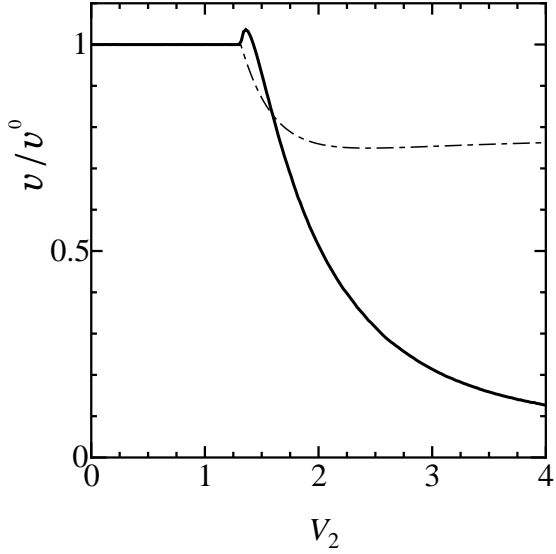}
 \vspace{-3mm}
\caption[]{
$V_2$ dependence of $v$, for $U=4$, $V=0$ and $\td=0$. 
The dot-dashed curve denotes the previous result obtained by 
a path integral method.
\cite{TomioPRB}
}
\end{center}
\end{figure}
%

 The spin-wave velocity $v$ corresponding to that in Fig.~7 
  is shown as a function of $V_2$ in Fig.~9 (solid curve). 
  The dot-dashed curve denotes the result obtained 
    previously by the path integral method. 
With increasing $V_2 ( > \V2c )$, the solid curve  
  decreases to zero rapidly while 
  the dot-dashed curve  tends to a finite value 
   ($v \rightarrow  t$) for large $V_2$. 
\cite{TomioPRB}
 Thus  the  present RPA calculation  indicates 
    the importance of 
 the couplings to  long-wavelength spin fluctuations,
 which has been  neglected 
 in the previous  path integral method.
The behavior in the limit of large $V_2$ will be discussed 
in the next section.

\section{Discussions}
\setcounter{equation}{0}

%
 We have examined spin-wave modes in the presence of  charge 
  ordering, which exhibits a clear  effect of suppressing 
   the  spin-wave velocity  even without dimerization ($\td = 0$).
 Since the dimerization enhances the effect of the repulsive interaction, 
  one may  expect further suppression of the spin-wave velocity 
   by adding the dimerization. 
 In this section, we discuss  
 such an effect of dimerization ($\td \not= 0$)
  together with  the limiting behavior  of large $V$ or $V_2$. 

First, we plot the velocity $v$ as a function of $V$ in Fig.~10, 
 with the fixed $\td=0$, 0.1, 0.3 and 0.5 ($U=4$ and $V_2=0$).  
The dimerization $\td$ has an effect of reducing  the velocity, 
which is qualitatively consistent with 
 the result of the path integral method.
\cite{TomioPRB} 
 The $V$ dependence of $v$ in the region II of Fig.~1   
  can be  fitted by a dotted curve, which 
is proportional to 
  $(t^2-\td^2)/(UV^2)$.
 We analyze such a case by noting that the ground state with  
   the spin and charge modulation  given by  
$(0, \uparrow, 0, \downarrow,0)$ and  $(0,1,0,1)$  
  already 
 shows up for $V \gsim 3$ even if $\td \neq 0$. 
In this state II, 
  the antiferromagnetic exchange coupling between 
 the up and down spins is induced by the process of 
 the fourth order of the transfers $t \pm \td$ leading 
   the effective Hamiltonian    as  
{%
\setcounter{enumi}{\value{equation}}
\addtocounter{enumi}{1}
\setcounter{equation}{0} 
\renewcommand{\theequation}{\arabic{section}.\theenumi\alph{equation}}%
\begin{eqnarray} \label{Heff2}
 H^{\rm eff}_{\rm I\!I} &\sim& J_{\rm I\!I} \sum^{N/2}_{j=1} 
 {\mib S}_{2j} \cdot {\mib S}_{2j+2} \virg \\
\label{J2}
 J_{\rm I\!I} &=& \frac{a_{\rm I\!I}(t^2-\td^2)^2}{UV^2} \virg
\end{eqnarray}
\setcounter{equation}{\value{enumi}}}%
where 
 $ a_{\rm I\!I} = 12$ and 
$
{\mib S}_{j}=(S^x_{j},S^y_{j},S^z_{j})
$ 
 with $S_j^\alpha= \psi^\dagger_{j} \sigma_{\alpha} \psi_{j}/2$
$
(\psi^\dagger_{j}=(C^{\dagger}_{j\uparrow},C^{\dagger}_{j\downarrow}))
$.  
This expression of $J_{\rm I\!I}$ with $a_{\rm I\!I} = 6$
has been obtained  by 
Mori and Yonemitsu.
\cite{Mori}  
 From eq.~(\ref{Heff2}), 
 the spin-wave dispersion is given by 
     $\omega(q)=J_{\rm I\!I} \sin(2q)$ 
  and then  the spin-wave velocity is given by 
           $v=2J_{\rm I\!I}$. 
 By comparing $v=2J_{\rm I\!I}$ with 
  the velocity calculated from eq.~(\ref{velocity}) 
in the inset of  Fig.~10,
     we found $a_{\rm I\!I} \simeq 1.6$. 
Such a value of $a_{\rm I\!I}$
 differs appreciably from that of the effective Hamiltonian 
 although one obtains a good coincidence between them 
 for the case at half-filling.
\cite{Tanemura_half}   
It is not yet clear at present why it is complicated to estimate  
 quantitatively by the  effective Hamiltonian at quarter-filling. 
%
\begin{figure}[tb]
\begin{center}
 \vspace{2mm}
 \leavevmode
 \epsfysize=7.5cm\epsfbox{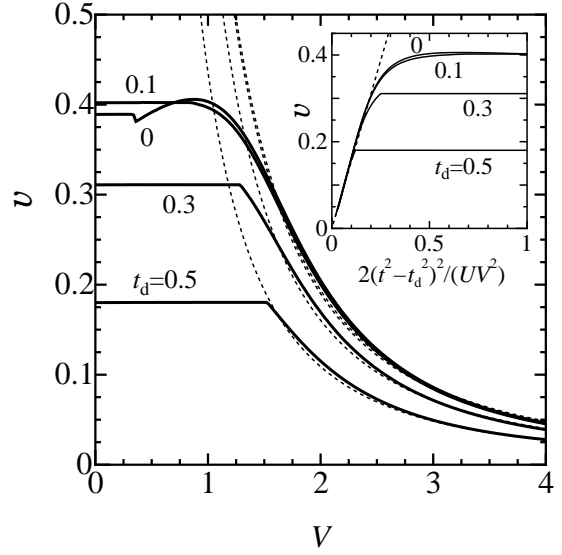}
 \vspace{-3mm}
\caption[]{
$V$ dependence of $v$, for $U=4$, $V_2=0$ and 
the fixed $\td=0$, 0.1, 0.3 and 0.5. 
The dotted curves represent 
 $v = 2J_{\rm I\!I} = 2 a_{\rm I\!I} (t^2-{\td}^2)^2/(UV^2)$, 
where $a_{\rm I\!I} \simeq 1.6$  is estimated from the inset. 
}
\end{center}
\end{figure}
%
%
\begin{figure}[tb]
\begin{center}
 \vspace{2mm}
 \leavevmode
 \epsfysize=7.5cm\epsfbox{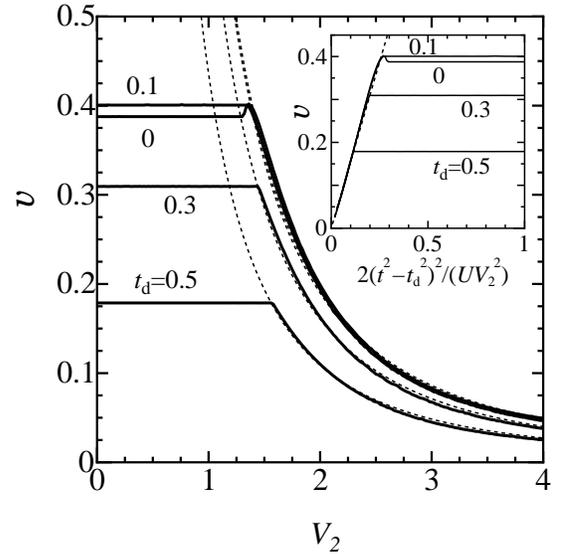}
 \vspace{-3mm}
\caption[]{
$V_2$ dependence of $v$, for $U=4$, $V=0$ and 
the fixed $\td=0$, 0.1, 0.3 and 0.5. 
The dotted curves represent 
$v=2\sqrt{{J}_{\rm I\!I\!I}{J}'_{\rm I\!I\!I}}
 \simeq 1.6 \times 2 (t^2-{\td}^2)^2/(UV_2^2)$, 
which is estimated from the inset. 
}
\end{center}
\end{figure}
%

Next, we examine  $V_2$ dependence of $v$ with some choices of $\td$, 
 which is shown  in Fig.~11. With increasing $\td$, 
  $v$ is suppressed monotonically in the region III (Fig.~1). 
 For arbitrary value of $\td$  in the  region III, 
 $v$  can be well fitted  by a dotted curve 
  which is proportional to $(t^2-\td^2)^2/(U V_2^2)$.  
 Such a behavior can be also analyzed 
  in terms of the following effective Hamiltonian. 
 In this case,  the spin and charge modulation in the ground state  
 is  given by   
$(0, \uparrow, \downarrow, 0, 0, \uparrow, \downarrow,0)$ 
and $(0,1,1,0,0,1,1,0)$, respectively. 
 The exchange coupling between the spins of 
  $(\uparrow,\downarrow)$  
   mainly comes from  the second  process 
  of  the transfer, while that between  the spins of 
$(\downarrow,0,0,\uparrow)$ comes from the sixth order process. 
Thus, the effective Hamiltonian for large $V_2$ 
 can be expressed as the Heisenberg model with the alternating interactions, 
 i.e., 
{%
\setcounter{enumi}{\value{equation}}
\addtocounter{enumi}{1}
\setcounter{equation}{0} 
\renewcommand{\theequation}{\arabic{section}.\theenumi\alph{equation}}%
\begin{eqnarray} \label{Heff3}
 H^{\rm eff}_{\rm I\!I\!I} &\sim&  \sum^{N/4}_{j=1} 
\left(
  J_{\rm I\!I\!I} {\mib S}_{4j-2} \cdot {\mib S}_{4j-1} 
 + J'_{\rm I\!I\!I} {\mib S}_{4j-1} \cdot {\mib S}_{4j+2} 
\right) \virg
\nonumber \\ \\
\label{J3}
  J_{\rm I\!I\!I} &=& \frac{ a_{\rm I\!I\!I} (t-\td)^2}{U}
\!\!\virg \\
\label{J3d}
  J'_{\rm I\!I\!I} &=& 
   \frac{a_{\rm I\!I\!I}' (t+\td)^2(t^2-\td^2)^2}{UV_2^4}
\virg 
\end{eqnarray} 
\setcounter{equation}{\value{enumi}}}%
 where $a_{\rm I\!I\!I} = 4$ and $a_{\rm I\!I\!I}' =1$.
 In this case,   
one can find the expression of the  spin-wave dispersion 
 (Appendix) as , 
$\omega(q) = \sqrt{J_{\rm I\!I\!I}J'_{\rm I\!I\!I}}\sin (2q)$, 
where  
$\sqrt{J_{\rm I\!I\!I}J'_{\rm I\!I\!I}} =  
 \sqrt{a_{\rm I\!I\!I} a_{\rm I\!I\!I}'} (t^2-\td^2)^2/(UV_2^2)$.
By comparing  
$v=2\sqrt{J_{\rm I\!I\!I}J'_{\rm I\!I\!I}}$ with 
 the numerical result in the inset of Fig.~11,
  we  find $\sqrt {a_{\rm I\!I\!I} a_{\rm I\!I\!I}'} \simeq 1.6$, which 
 is compatible with that of the effective Hamiltonian, i.e., 2. 

In summary, 
 we have calculated  the spin-wave excitations for 
 a one-dimensional model with a quarter-filled band 
 to investigate 
    the effects of the nearest ($V$) and 
 next-nearest-neighbor ($V_2$) 
 interactions on them.
 From the dispersion relations for spin-wave modes 
 in the coexistent state of SDW and CDW, we found that 
 the spectrum of acoustic mode and  then 
 the spin-wave velocity are reduced by  CDW   
  (induced by $V$ or $V_2$)
 and that the spin-wave   
 in the coexistent state  for large $V$ or $V_2$ 
 could  be described by spin 1/2 antiferromagnetic Heisenberg models with 
   the effective exchange coupling.

\section*{ Acknowledgments }

The author (Y.K.) would like to express his appreciation 
 to Professor Y. Kuroda for  stimulating discussions 
and the warm hospitality during his stay at Nagoya University.  
 This work was done at Nagoya University,  
where Y.K. has been financially supported by  
a grant from Hokkaido Tokai University.

\appendix
\section{Spin-Wave of the Heisenberg Antiferromagnets}
\setcounter{equation}{0}

By taking four lattice sites as a unit cell and 
 introducing the sublattice $A$ and $B$, 
 the Hamiltonian of eqs. (4.1a) and (4.2a) can be rewritten as 
\begin{eqnarray} \label{Heff}
 H^{\rm eff} &=&  \sum^{N/4}_{l=1}   
\left(
  J {\mib S}_{B,l} \cdot {\mib S}_{A,l} 
 + J' {\mib S}_{B,l} \cdot {\mib S}_{A,l+1} 
\right)
\virg
\end{eqnarray} 
where $J=J'=J_{\rm I\!I}$ for eq. (\ref{Heff2}),  
 and $J=J_{\rm I\!I\!I}$ and $J'=J'_{\rm I\!I\!I}$ 
for eq. (\ref{Heff3}). 
 By using the Holstein-Primakoff transformation, 
\begin{eqnarray} 
\label{HPAz}
 S^z_{A,l} &=&  1/2-a^\dagger_l a_l   \virg \\
\label{HPA+}
 S^+_{A,l} &=& S^x_{A,l}+\i S^y_{A,l}
= (1-a^\dagger_l a_l)^{1/2} a_l    \virg \\
\label{HPA-}
 S^-_{A,l} &=& S^x_{A,l}-\i S^y_{A,l}
= a^\dagger_l (1-a^\dagger_l a_l)^{1/2}     \virg \\
\label{HPBz}
 S^z_{B,l} &=&  -1/2+b^\dagger_l b_l   \virg \\
\label{HPB+}
 S^+_{B,l} &=& S^x_{B,l}+\i S^y_{B,l}
= b^\dagger_l (1-b^\dagger_l b_l)^{1/2}    \virg \\
\label{HPB-}
 S^-_{B,l} &=& S^x_{B,l}-\i S^y_{B,l}
= (1-b^\dagger_l b_l)^{1/2} b_l     \virg
\end{eqnarray} 
eq. (\ref{Heff}) up to the second order 
 with respect to  $a_q$ and $b_q$ 
is given by 
\begin{eqnarray} \label{Heff-a}
 H^{\rm eff} &\simeq& \frac{1}{2}\sum_{q}   
\left[ 
(J + J') \left(b^\dagger_q b_q + a^\dagger_q a_q \right)
\right. \nonumber \\
& & \left. \hspace{-8mm} {}
+ (J + J'\e^{-\i4q}) b^\dagger_q a^\dagger_q 
+ (J + J'\e^{\i4q}) b_q a_q \right]
\virg
\end{eqnarray} 
where  the constant term is subtracted and 
 the boson operators $a^\dagger_q$ and $b^\dagger_q$ are 
  given by 
\begin{eqnarray} 
\label{aq}
 a^\dagger_q &=& (N/4)^{-1/2} \sum_{l} \e^{\i q R_l}\: a^\dagger_l 
\virg \\
\label{bq}
 b^\dagger_q &=& (N/4)^{-1/2} \sum_{l} \e^{-\i q R_l}\: b^\dagger_l 
\point
\end{eqnarray} 
By applying the Bogoliubov transformation to eq.(\ref{Heff-a}),  
 the spin-wave dispersion, $\omega(q)$, is obtained from  
\begin{eqnarray} \label{det-w}
\det \left( 
 \begin{array}{cc}
 \omega(q)-(J+J')/2  &   - (J+J'\e^{\i4q})/2       \\
 (J+J'\e^{-\i4q})/2  &   \omega(q)+(J+J')/2   
\end{array}
\right) &=& 0 
\point \nonumber \\
\end{eqnarray} 
Equation  eq.~(\ref{det-w}) leads to   the spin-wave dispersion 
 for the Hamiltonian (\ref{Heff}) as 
\begin{eqnarray} \label{omega-q}
 \omega(q) &=& \sqrt{JJ'}\sin(2q) \virg
\end{eqnarray} 
where 
$\omega(q) = J_{\rm I\!I}\sin(2q)$ 
for the Hamiltonian (\ref{Heff2}) and 
$\omega(q) = \sqrt{J_{\rm I\!I\!I} J'_{\rm I\!I\!I}}\sin(2q)$ 
for the Hamiltonian (\ref{Heff3}).



\end{document}